# Exploring Educational Equity: A Machine Learning Approach to Unravel Achievement Disparities in Georgia

Yichen Ma, Dr. Dima Nazzal

**Abstract**
The COVID-19 pandemic has significantly exacerbated existing educational disparities in Georgia's K-12 system, particularly in terms of racial and ethnic achievement gaps. Utilizing machine learning methods, the study conducts a comprehensive analysis of student achievement rates across different demographics, regions, and subjects. The findings highlight a significant decline in proficiency in English and Math during the pandemic, with a noticeable contraction in score distribution and a greater impact on economically disadvantaged and Black students. Socio-economic status, as represented by the Directly Certified Percentage – the percentage of students eligible for free lunch, emerges as the most crucial factor, with additional insights drawn from faculty resources such as teacher salaries and expenditure on instruction. The study also identifies disparities in achievement rates between urban and rural settings, as well as variations across counties, underscoring the influence of geographical and socio-economic factors. The data suggests that targeted interventions and resource allocation, particularly in schools with higher percentages of economically disadvantaged students, are essential for mitigating educational disparities.

**1. Introduction**
In the wake of the COVID-19 pandemic, the American education system has faced monumental challenges. The pandemic disrupted the learning experience for millions of students and exposed deep-rooted inequalities [1] [2]. Among the most alarming revelations is the exacerbated racial and ethnic disparities in student achievement. National test results in 2022 indicated a significant decline in math and reading performance among 9-year-olds, the lowest in two decades [3]. This decline was evident across all races and income levels, amplifying the pre-existing disparities. In the aftermath of the pandemic, students experienced a greater decline in math and science compared to other disciplines and have yet to recover from the initial educational setbacks even after a year [2]. The state of Georgia, in particular, witnessed its vulnerable and at-risk student populations bear the brunt of the pandemic's consequences [4].

The achievement gap is not a new phenomenon; it has been a persistent issue within the American education system. However, the current crisis has renewed the urgency to address this issue. The achievement gap refers to the differences in educational performance and outcomes among various racial and ethnic groups. In Georgia, for example, White students demonstrated a significantly higher average achievement rate of 66.99%, compared to 39.88% for Black students in 2022. The achievement gap has far-reaching implications, as it not only affects individual students but also has broader socio-economic consequences, such as reduced earning potential and increased likelihood of unemployment [5]. Simultaneously, the achievement rate serves as an essential metric in evaluating the education system's efficacy. The achievement rate is a measure of the percentage of students who meet or exceed the standard academic proficiency level. It reflects the collective performance of students in a given group or area, such as a school, district, or even a state. In the context of Georgia, despite overall declines, certain schools and districts have maintained a relatively high achievement rate, suggesting the presence of effective strategies or factors contributing positively to student performance.

The relationship between the achievement rate and the achievement gap is intricate. A high achievement rate does not necessarily imply a small achievement gap, especially in diverse communities where different racial and ethnic groups may exhibit varying levels of achievement. Conversely, reducing the achievement gap can lead to an increase in the overall achievement rate if underperforming groups improve their performance.

This research paper aims to employ machine learning methods to identify effective strategies to bridge racial and ethnic disparities in the achievement gap among K-12 students in Georgia. By predicting achievement rates and disparities, identifying factors that contribute significantly to these disparities, this study intends to propose policies and strategies that can be implemented to mitigate the achievement gap. Moreover, by understanding the interplay between the achievement rate and the achievement gap, this study strives to offer insights into elevating overall student performance while ensuring equity. This endeavor is crucial in shaping an education system that provides equal opportunities and fosters the holistic development of all students, regardless of their racial or ethnic background. Additionally, this study analyzes the impact of Covid-19 on achievement rates, highlighting disparities across racial groups, economically disadvantaged students, and differences between urban and rural settings, among other aspects.

**2. Background**

Understanding the nuances of educational achievement in the context of racial and ethnic disparities has long been a focal point of academic research, policy formulation, and educational interventions. To provide a comprehensive perspective, this background section is structured into three main segments: a thorough analysis of the achievement gap and achievement rate, a review of interventions and the unique role of machine learning, and an examination of the relationship with prior research, while also considering ethical implications.

## 2.1. Understanding the Achievement Gap and Achievement Rate

The achievement gap is not merely a statistic but represents a multifaceted and profound challenge, underpinned by disparities in academic performance across different racial and ethnic groups.

*Socio-Economic Factors*: Extensive research has documented the role of socio-economic status in contributing to this gap [6]. The very institutions meant to provide educational opportunities can inadvertently perpetuate disparities, especially in the context of socio-economic factors like poverty. As poverty directly influences student achievements through the quality of education, the role of schooling systems in either mitigating or exacerbating these effects is pivotal [7]. Students from higher socio-economic backgrounds usually exhibit better academic results, while those from lower socio-economic backgrounds often experience less preparation for school and thus poorer outcomes [8]. The complex interplay between socio-economic status, family environment, and educational achievements has led to a continuous debate on devising strategies to mitigate these effects [9]. Furthermore, a meta-analysis of correlational evidence has demonstrated that both risk and protective factors significantly influence academic achievement, suggesting targeted interventions could be developed to address these disparities [10].

*Educational Resources and School Funding*: Disparities in educational resources further exacerbate the achievement gap. Schools with a higher concentration of minority students frequently face funding challenges, which translates into reduced availability of essential resources [11]. Funding on education in the United States is predominantly a responsibility of state and local entities, with the federal government contributing about 8% of the total funds, which includes programs beyond the Department of Education like Head Start and the School Lunch program [12]. However, local funding, primarily from property taxes, along with state and private contributions, account for the remaining 92% of school finances. This funding structure leads to disparities, as schools in less affluent areas receive fewer funds per student, causing a "financial segregation." Consequently, minority students in lower-valued property areas often face disproportionate funding challenges. [13]. This scenario perpetuates an unequal playing field, where the academic opportunities available to students are directly linked to their geographical location and racial or ethnic background.

*Parental Involvement and Systemic Issues*: Parental involvement has been identified as a key factor influencing the achievement gap, with active engagement often correlating with improved outcomes [14]. Children with parents who have lower educational backgrounds often exhibit lower educational achievement compared to those with highly educated parents [15]. This disparity, as highlighted by Hosokawa and Katsura [16], can be attributed to family processes that influence a child's adjustment from preschool through early elementary school. In their longitudinal study, they emphasize the role of social competence, suggesting that socio-economic status interacts with family dynamics, subsequently shaping a child's social skills and overall school adjustment. Parents with advanced education levels typically offer more socio-economic resources, which, combined with effective family processes, can bolster a child's social competence and positively impact their academic performance. Alongside this, the shadow of historical discrimination, biases in standardized testing, and systemic issues continue to haunt the American education system, leading to the disengagement of certain groups, such as Black students [17].

## 2.2 Interventions and the Role of Machine Learning

In response to the achievement gap, various interventions have been suggested, including policies focusing on early childhood education [18], equitable school funding, parental involvement [19], and culturally responsive teaching. However, while these interventions have shown promise, there remains a gap in the application of advanced quantitative methodologies, such as machine learning, to analyze and address this issue. Machine learning techniques offer the advantage of handling vast amounts of data and identifying complex, non-linear relationships that might be overlooked by traditional statistical methods. By harnessing these capabilities, we can advance our understanding of the intricate web of confounding factors that contribute to the achievement gap. Furthermore, machine learning can help in uncovering hidden patterns and providing insights into potential actionable interventions. This approach ensures that the interventions are data-driven and tailored to address the specific needs of the marginalized minority groups. This research aims to fill this gap by employing machine learning techniques to predict the disparities among marginalized minority groups in Georgia based on various school features. The study also incorporates the analysis of feature weights to ascertain the most critical and sensitive factors contributing to the observed disparities. Through this, the research aims to propose targeted and effective intervention strategies.

## 2.3 Advancing Prior Research

Two recent studies provide a useful starting point for this research. The first study explored the use of machine learning techniques to predict student achievement in higher education. The authors found that different algorithms, such as Decision

Trees, Random Forests, Logistic Regression, K-Nearest Neighbors, Gradient Boosting, AdaBoost, Gaussian Naive Bayes, and Support Vector Machines, had varying degrees of success when dealing with socio-demographic data. The Random Forest algorithm was found to be the most effective, with an accuracy rate of about 89.39% [20]. In contrast, the second study focused on high school students, employing psychological tests and scales as a foundation for its predictions. It revealed that the Support Vector Machine (SVM) algorithm was most effective in this context [21].

In [22], the authors utilize regression and decision tree analysis with the CHAID algorithm to delve into how students' attributes and experiences influence their satisfaction levels. From their analysis of student-opinion data, the researchers identified that different aspects of a student's university experience predominantly affect three distinct measures of general satisfaction. Notably, these three satisfaction measures have unique predictors, emphasizing their non-interchangeability. Among the influential factors, academic experiences stand out, with faculty preparedness – a factor previously linked to student achievement – being a primary determinant of student satisfaction. Additionally, the study found that while social integration and pre-enrollment opinions play significant roles in determining satisfaction, campus services, facilities, and students' demographic attributes do not serve as significant predictors. Furthermore, the decision tree analysis highlighted that social integration notably affects the satisfaction of students who exhibit lower academic engagement.

[23] further employs advanced machine learning techniques to analyze the factors influencing students' PISA 2015 test scores across nine countries. Their primary objectives are to identify the student attributes correlated with these scores and discern the school characteristics linked to school value-added measures. Crucially, the study aims to uncover non-linear relationships between these factors and test scores and explore how school-level factors interact to influence results. Adopting a two-stage approach with tree-based methods, the research first employs multilevel regression trees to estimate school value-added, followed by leveraging regression trees and boosting in the second stage to associate the estimated value-added to school-level variables. This method offers a refined understanding of the diverse educational production functions in different nations.

Beyond the standard machine learning approaches, this research deploys sophisticated techniques like Neural Networks and two distinct Deep Learning models, each meticulously configured to capture intricate patterns in the data. A significant enhancement is the use of SHAP (SHapley Additive exPlanations) values, providing a nuanced understanding of feature importance in the models, a facet often overlooked in conventional studies. The dual approach of using both traditional neural networks and deep learning models ensures an exhaustive exploration of the dataset. Additionally, the integration of geospatial visualization and the meticulous hyperparameter tuning through Randomized Search with cross-validation further distinguishes our methodology. This robust approach not only identifies the determinants of student achievement but also provides actionable insights with a high degree of interpretability, making it exceptionally valuable for stakeholders in the educational sector.

In summary, this research builds upon existing literature to provide a comprehensive and nuanced analysis of the achievement gap and achievement rate in Georgia. By combining diverse data types and methodologies, and considering ethical dimensions, it aims to make a meaningful contribution to the field.

## 3. Methods
### 3.1. Data
The data are collected from the Georgia Department of Education and the Governor's Office of Student Achievement [24] [25]. We have included all the related data we can find. The achievement rate or the achievement score utilizes weights based on achievement level and is calculated by $Beginner\ Learner \times 0 + Developing\ Learner \times 0.5 + Proficient\ Learner \times 1 + Distinguished\ Learner \times 1.5$. Content Mastery indicators address whether students are achieving at the level necessary to be prepared for the next grade, college, or career. It includes achievement scores across four subjects based on student performance on the Georgia Milestones Assessment System and the Georgia Alternate Assessment (GAA) 2.0. Table 1 lists the variables included in the data sets and the type of the variable.

### 3.2. Machine Learning Models
The initial phase of the study encompasses pre-processing, which is vital for refining data quality and includes feature selection using the *SelectKBest* class from scikit-learn. By employing mutual information as a criterion, relevant features indicative of the target variable are isolated. The dataset is subsequently divided into training and testing subsets, a standard practice ensuring the evaluation of the model on new data. Additionally, feature scaling via the *StandardScaler* method is incorporated, essential for algorithms that are sensitive to feature magnitudes. The modeling stage involves the utilization of several machine learning algorithms, each chosen for specific strengths. Logistic Regression is chosen for its simplicity and ease of interpretation. Decision Trees, though non-parametric and efficient in capturing non-linear relationships, tend to overfit, and to address this, hyperparameters like maximum depth and minimum samples for internal node splitting are tuned. Recognizing the limitations of single trees, Random Forest, an ensemble algorithm comprising multiple decision trees, is employed. In contrast to the independent construction of trees in Random Forest, the study also incorporates XGBoost, an ensemble method that builds trees sequentially, ensuring that each subsequent tree addresses the errors of its predecessors.

| Variable | Type |
|---|---|
| *Pre-Pandemic All Math Achievement Rate* | Quantitative |
| *Post Pandemic All Math Achievement Rate* | Quantitative |
| *Absent 0-5 Days Percentage* | Quantitative |
| *Absent 6-15 Days Percentage* | Quantitative |
| *Absent 15+ Days Percentage* | Quantitative |
| *Average Annual Salaries - Administration* | Quantitative |
| *Average Annual Salaries - Teachers* | Quantitative |
| *Average Annual Salaries – Support Personnel* | Quantitative |
| *Number of Bachelor Degree Teachers* | Quantitative |
| *Number of Master Degree Teachers* | Quantitative |
| *Number of Specialist Degree Teachers* | Quantitative |
| *Number of PHD Degree Teachers* | Quantitative |
| *Number of Certified Teachers* | Quantitative |
| *Total Students Enrolled* | Quantitative |
| *Teacher-Student Ratio* | Quantitative |
| *White Student Percentage* | Quantitative |
| *Black Student Percentage* | Quantitative |
| *Economically Disadvantaged Student Percentage* | Quantitative |
| *Students with Disability Percentage* | Quantitative |
| *English Learners Percentage* | Quantitative |
| *Directly Certified Percentage* | Quantitative |
| *Expenditure Instruction* | Quantitative |
| *Expenditure School Administration* | Quantitative |
| *Expenditure Pupil Services* | Quantitative |
| *Per-Pupil Expenditure* | Quantitative |
| *Per-Pupil Expenditure Federal* | Quantitative |
| *Rate of Entries and Withdraws* | Quantitative |
| *Percentage of Gifted Students* | Quantitative |
| *Urban/Rural Classification of the School* | Categorical |
| *Grade Level Classification* | Categorical |
| *School District Classification* | Categorical |
| **Derived Variable** | **Type** |
| *Expenditure Instruction Normalized* | Quantitative |
| *Post Grad Percentage* | Quantitative |
| *2019 - 2021 Achievement Growth Rate* | Quantitative |

*Table 1. Data Definition*

Additionally, the study leverages the capabilities of neural networks through Multi-Layer Perceptron (MLP), which processes data in complex patterns across multiple layers of neurons. Support Vector Machines (SVM) are employed for their adeptness in handling both linear and non-linear data by finding optimal hyperplanes for class separation. Furthermore, the K-Nearest Neighbors algorithm (KNN) finds utility in classifying data points by considering the majority class amongst its neighbors. Ensemble methods, including Voting, Stacking, and Bagging, are also an integral part of the study. These methods are employed to combine the predictions of multiple machine learning models with the aim of improving the generalization error and robustness. The performance of each model is meticulously evaluated using several metrics, including training and testing accuracy, mean squared error (MSE), and R-squared statistics.

A distinguishing feature of this study is the analysis of feature importance, particularly for models that support such extraction. This interpretability is important, especially for policymakers and educational analysts looking for insights that can translate into actionable interventions. Moreover, the study pioneers the use of geospatial visualization using *GeoJSON*, providing spatial understanding and depth, which is essential in observing spatial patterns and correlations with achievement rates.

**3.3. Advanced Models**
In this section, three distinct models are discussed, which are utilized to predict student achievement rates based on various features. These models consist of one traditional neural network and two deep learning models. These models are selected to efficiently capture the underlying patterns in the dataset and to make predictions with a high degree of accuracy.

*Neural Network Model*
The first model employed is a Neural Network. Neural Networks are a subset of machine learning, essentially composed of neurons arranged in layers. These neurons learn by adjusting the weights applied to the input as well as their bias. A simple neural network consists of an input layer, one or more hidden layers, and an output layer. Neural Networks have been widely used in solving regression and classification problems. In this approach, the neural network is configured with an input layer, three hidden layers, and an output layer (Figure 1). The size of the input layer is determined by the number of features in the dataset. Hidden layers are used to capture the complexity in the data. The first hidden layer has 128 neurons with a Rectified Linear Unit (ReLU) activation function, and the second and third hidden layer has 64, 32 neurons with the same activation function. The output layer has a single neuron with a linear activation function, as this is a regression problem. Dropout regularization is also incorporated, which randomly sets a fraction of input units to 0 at each update during training time, helping to prevent overfitting. This model uses the Adam optimizer, which adapts the learning rate during training, and the mean squared error as the loss function since this is a regression problem.

One of the advantages of using a neural network is its ability to model complex relationships between inputs and its capacity to capture both linear and non-linear relationships in the data. However, it requires a significant amount of data, can be computationally expensive, and is sensitive to feature scaling. Moving forward, deep learning models are used. Deep learning is an extension of neural networks with many hidden layers, which allows the model to learn increasingly abstract features from the raw input. They have been particularly effective in fields such as image and speech recognition.

*Deep Learning Models*
The first deep learning model contains more layers and neurons than the simple neural network. The structure comprises an input layer followed by three hidden layers and an output layer. Specifically, the first hidden layer has 128 neurons, the second hidden layer has 64 neurons, and the third hidden layer has 32 neurons. All the hidden layers use the ReLU activation function. Dropout regularization is used after the first and second hidden layers. Like the neural network model, this model uses the Adam optimizer and mean squared error as the loss function. The second deep learning model is similar to the first deep learning model but has more neurons in the first hidden layer, 256 neurons to be exact, and incorporates additional dropout layers.

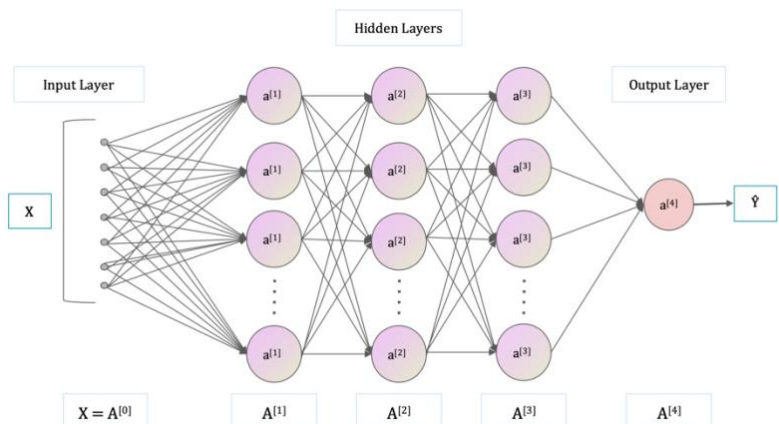

*Figure 1. Neural Network Structure*

To visualize the impact of each feature on the models' output, SHAP (SHapley Additive exPlanations) values are used. SHAP is a unified measure of feature importance and is based on cooperative game theory. SHAP values not only show the importance of each feature but also about the direction of the relationship with the target variable (positive or negative). In conclusion, through the use of neural networks and deep learning models, complex relationships in the data can be captured. These models have been tuned for better performance and their results can be interpreted through R-squared scores and SHAP values.

## 4. Results
### 4.1. Achievement Rate
In a comparative analysis of student achievement rates among the 2019, 2021, and 2022 school years, there was a significant decline in both Math and English proficiency in 2021, followed by a partial recovery in 2022. For Math, the mean achievement rate decreased from 63.16% in 2019 to 46.20% in 2021, before increasing to 57.31% in 2022 (Figure 2). Similarly, English scores experienced a drop from a mean of 63.23% in 2019 to 50.23% in 2021, with a subsequent rise to 56.58% in 2022 (Figure 3). Additionally, the data shows a shrinking of the interquartile range (IQR) in both subjects, indicating a tightening in score distribution. In 2021, the IQR for Math narrowed, as evidenced by a higher Q1 (39.93%) compared to 2019 (49.69%), and a lower Q3 (73.50% from 79.74%). The English scores also saw a tightened IQR with Q1 moving up to 42.70% in 2021 from 35.70% in 2019 and Q3 decreasing to 70.37% in 2022 from 78.36% in 2019. This suggests a decrease in average scores and a contraction in the distribution of scores, with fewer students achieving the upper quartiles in 2021 compared to 2019, and a slight dispersion in 2022. Henceforth, the Math achievement rate will serve as our primary metric for analysis due to its more pronounced variability compared to English.

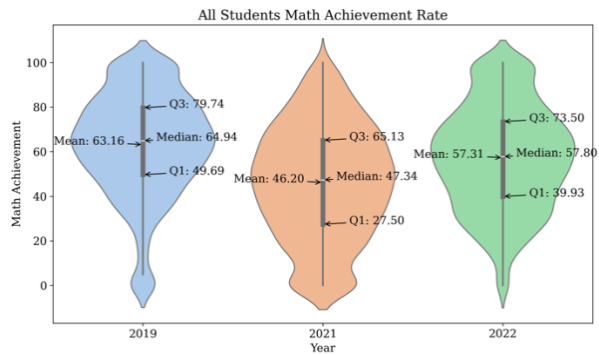

*Figure 2. All Students Math Achievement Rate for 2019, 2021, and 2022*

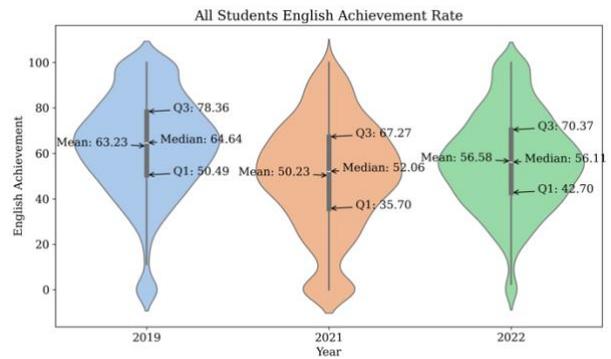

*Figure 3. All Students English Achievement Rate for 2019, 2021, and 2022*

The distributions of math achievement rates in 2021 across different student demographics underscore marked disparities. Economically disadvantaged students show a mean achievement rate of 50.05% with a median of 50.09%, while the general student body has a higher mean of 57.89% and a median of 57.91% (Figure 4). This indicates a persistent achievement gap between economically disadvantaged students and their peers. Furthermore, a comparison of achievement rates by race reveals a pronounced disparity: White students have a mean rate of 70.08% with a median of 73.85%, considerably higher than Black students, who have a mean of 48.57% and a median of 49.78% (Figure 5). The interquartile ranges also reflect this gap, with White students having a wider spread (Q1: 58.85, Q3: 87.85) compared to Black students (Q1: 36.26, Q3: 60.87), suggesting a broader distribution of achievement within the White student group.

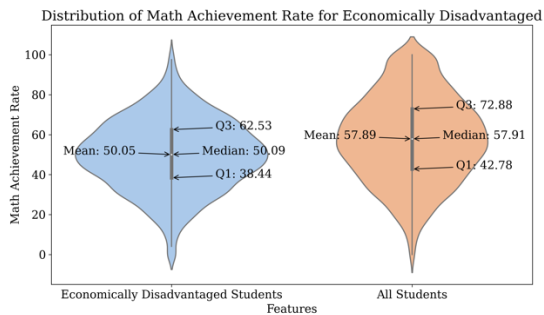

*Figure 4. Math Achievement Rate for Economy Classification*

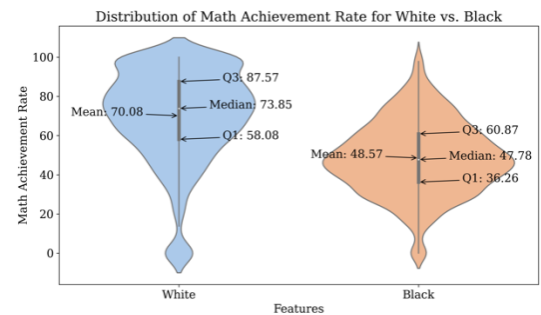

*Figure 5. Math Achievement Rate for Race Classification in 2021*

**4.1.1. Cross-County Comparison**

Georgia County Maps below indicates how the factors play a role across 159 counties in Georgia with yellow stars indicating the major cities – Atlanta, Savannah, Augusta, Athens, Macon, Valdosta, and Albany.

*Achievement Rate in 2019 and 2022*

In both 2019 and 2022, certain counties stand out for their consistent academic excellence. Specifically, counties Forsyth, Oconee, Fayette, Pierce, Camden, Bryan, and Lowndes have shown persistently high achievement rates in Math and English. The visual data presented in Figures 6 A-D shows this trend across the years.

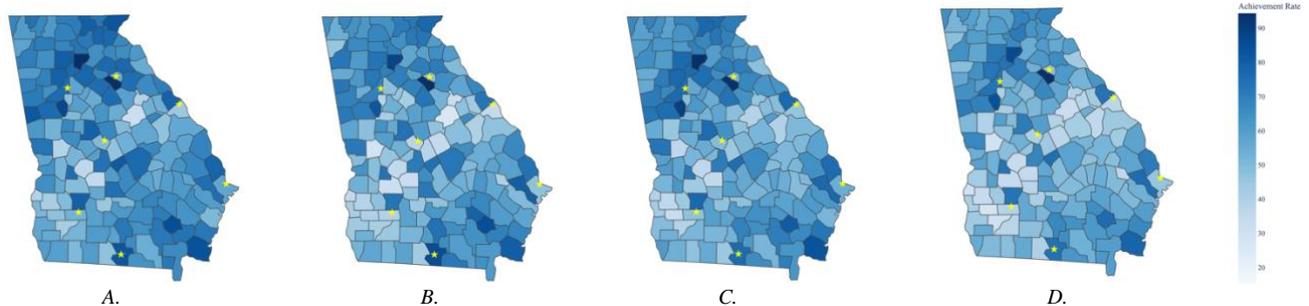

*Figure 6 A. Visualization of Math Achievement Rate in 2019 in Georgia at County Levels*
*Figure 6 B. Visualization of Math Achievement Rate in 2022 in Georgia at County Levels*
*Figure 6 C. Visualization of English Achievement Rate in 2019 in Georgia at County Levels*
*Figure 6 D. Visualization of English Achievement Rate in 2022 in Georgia at County Levels*

*Growth Rate from 2019 to 2021 and 2022*

During the pandemic, the growth rate in Math achievement across all students from 2019 to 2021 (Figure 7A) revealed that only five counties – Mclntosh, Pulaski, Lowndes, Echols, and Putnam in order from the highest growth to the lowest – demonstrated positive growth, suggesting that most counties saw a decline. This decline can be attributed to the pandemic-induced disruptions in traditional learning environments. In contrast, when assessing the English achievement growth rate (Figure 7C), a mere three counties – Fulton, McIntosh, and Clarke – reported positive progress. Notably, even though fewer counties registered growth in English, the average decline in English stood at 9.88%, while the average decline in English stood at 13.79%, possibly due to Math's inherent complexity compared to English.

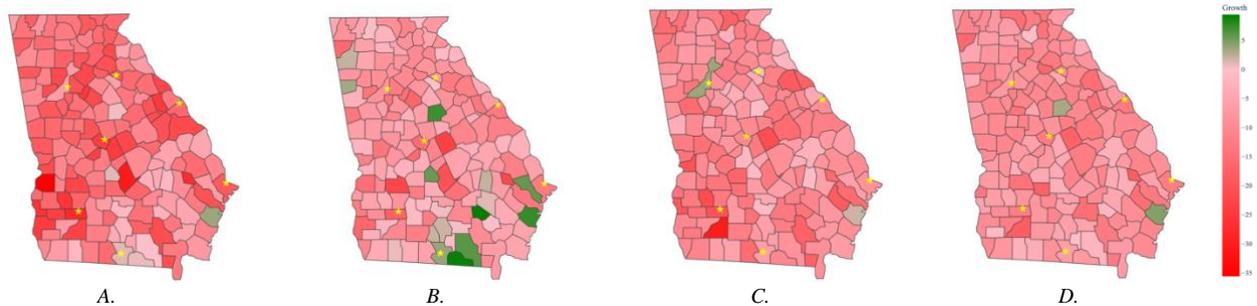

A.         B.         C.         D.

Figure 7 A. Visualization of Math Achievement Growth Rate (2019 to 2021) in Georgia at County Levels
Figure 7 B. Visualization of Math Achievement Growth Rate (2019 to 2022) in Georgia at County Levels
Figure 7 C. Visualization of English Achievement Growth Rate (2019 to 2021) in Georgia at County Levels
Figure 7 D. Visualization of English Achievement Growth Rate (2019 to 2022) in Georgia at County Levels

| *County* | *Overall English Ach. Rate Growth (%)* | *Urban/Rural* | *County* | *Overall English Ach. Rate Growth (%)* | *Urban/Rural* |
|---|---|---|---|---|---|
| *Fulton* | 3.66 | Urban | *Mitchell* | -30.41 | Rural |
| *Mclntosh* | 1.79 | Urban | *Terrell* | -21.24 | Urban |
| *Clarke* | 0.48 | Urban | *Wilkinson* | -20.91 | Rural |
| *Toombs* | -0.53 | Rural | *Chattahoochee* | -20.23 | Urban |
| *Putnam* | -0.69 | Rural | *Dougherty* | -20.08 | Urban |

Table 2. Top 5 and Bottom 5 Counties' Overall English Growth

| *County* | *Overall Math Ach. Rate Growth (%)* | *Urban/Rural* | *County* | *Overall Math Ach. Rate Growth (%)* | *Urban/Rural* |
|---|---|---|---|---|---|
| *Mclntosh* | 3.65 | Urban | *Stewart* | -35.53 | Rural |
| *Pulaski* | 1.08 | Urban | *Dodge* | -30.75 | Rural |
| *Lowndes* | 1.05 | Urban | *Dougherty* | -28.05 | Urban |
| *Echols* | 0.72 | Urban | *Evans* | -27.35 | Rural |
| *Putnam* | 0.46 | Rural | *Clay* | -26.26 | Rural |

Table 3. Top 5 and Bottom 5 Counties' Overall Math Growth

*Urban vs. Rural*
We employed the 2013 Rural-Urban Continuum Codes form a classification scheme that distinguishes metropolitan counties by the population size of their metro area, and nonmetropolitan counties by degree of urbanization and adjacency to a metro area. The official Office of Management and Budget (OMB) metro and nonmetro categories have been subdivided into three metro as Urban and six nonmetro categories as Rural. Each county in Georgia is assigned one of the 9 codes. Figure 8 shows the urban and rural classification at county levels.

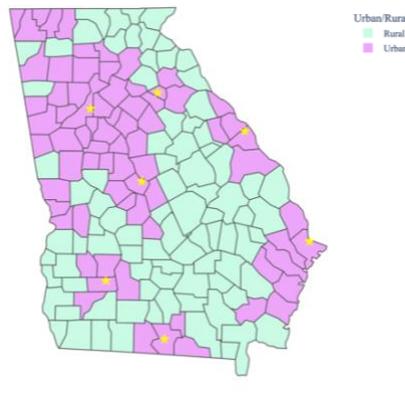

*Figure 8. Visualization of Urban/Rural Classification in Georgia at County Levels*

For Math achievement rates (Figure 8), in 2019, Georgia's rural counties exhibited a mean achievement rate of 62.06%, with the middle 50% of counties (from Q1 to Q3) having achievement scores between 52.54% and 75.84%. Comparatively, urban counties slightly outperformed and presented a mean achievement rate of 66.25%, with their middle 50% achieving scores from 51.98% to 82.98%. By 2021, the dynamics shifted. Rural counties recorded a reduced mean achievement rate of 50.34%, with their scores' central distribution between 37.00% and 64.79%. Meanwhile, urban counties saw a bigger drop with an average score of 47.53% with their middle 50% spanning from 29.07% to 66.96%. Notably, this data suggests that while urban counties had an advantage in 2019, by 2021, rural counties slightly edged ahead, reflecting the differential impacts of the pandemic on educational structures.

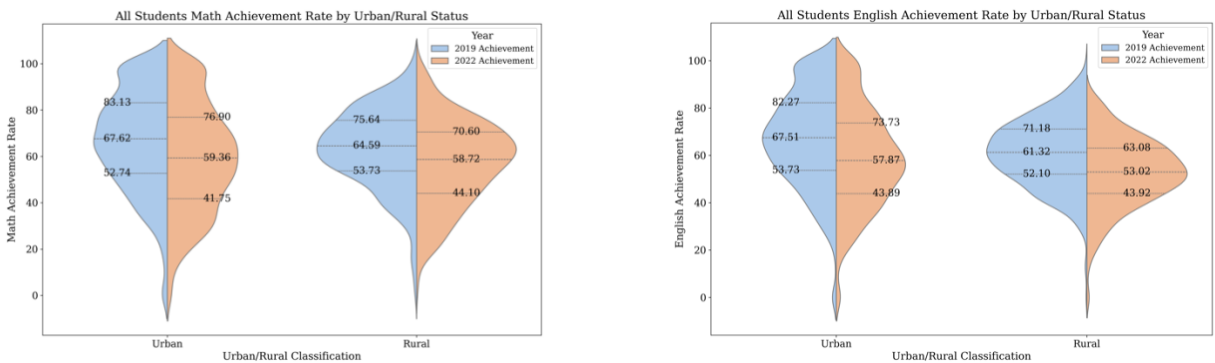

*Figure 9 (left). Math Achievement Rate for Urban and Rural Classifications in 2019 and 2022*
*Figure 10 (right). English Achievement Rate for Urban and Rural Classifications in 2019 and 2022*

Turning to English achievement rates (Figure 9), 2019 data depicts rural counties with a mean achievement of 59.78%, with their central 50% of scores oscillating between 50.90% and 71.16%. In contrast, urban counties during this period exhibited a more substantial mean rate of 66.52%, with their 50% core scores ranging between 53.10% and 81.74%. Transitioning to 2021, the figures reveal a decline for both categories. Rural counties' mean achievement descended to 49.90%, with their central scores between 41.03% and 61.24%. Concurrently, urban counties too reported a reduction, settling at a mean of 52.04% and a middle 50% distribution from 36.58% to 69.53%.

*Economically Disadvantaged Student Percentage*
Economically disadvantaged students' percentage varies by county and can be influenced by factors such as geographic location, urban or rural classification, and local economic conditions. In general, counties with higher levels of poverty tend to have a higher percentage of economically disadvantaged students in their K-12 public schools.

In terms of geographic distribution, economically disadvantaged students are more concentrated in certain areas of the state. Observing the pattern of the Economically Disadvantaged student percentage (Figure 11), there's a clear correlation with a county's urbanization level. For example, the southern region of Georgia has a higher percentage of economically disadvantaged students compared to the northern region. The top 10 counties with the highest percentage of economically disadvantaged students were all rural and located in the southern part of the state. These counties include Clay, Taliaferro, and Randolph counties, which all had more than 80% of their students classified as economically disadvantaged. This is likely due to factors such as a higher prevalence of poverty and limited economic opportunities in certain areas of the state. On the other hand, urban counties in Georgia may also have a significant percentage of economically disadvantaged students, particularly those with a high population density and high poverty rates. For example, Fulton County, which includes the city of Atlanta, had a 42% economically disadvantaged student percentage in 2021.

In terms of achievement rates, economically disadvantaged students in rural schools performed better than their counterparts in urban schools, but the pandemic may have contributed to a decline in overall proficiency rates for economically disadvantaged students in both urban and rural schools. In 2019, rural schools displayed a median Math Achievement Rate of 58.93%, with a first quartile (Q1) at 48.29 and a third quartile (Q3) reaching 69.30. Urban schools in the same year posted a slightly lower median of 53.49%, where the interquartile range (from Q1 at 40.19 to Q3 at 68.16) suggests a broader distribution of outcomes. However, the onset of the pandemic appears to have worsen these rates across both settings, with rural schools witnessing a median drop to 47.56% by 2021, and urban schools experiencing a more pronounced median decline to 34.28%. The interquartile ranges contracted in both scenarios, indicating a tightening distribution of achievement rates.

By 2022, a recovery was evident, especially in urban schools, where the median increased about 10 percent from 34.28% to 44.55%. Rural schools, while also slightly recovered, maintained a marginally higher median of 52.92%, suggesting a degree of resilience in these environments. The reasons for the better performance of economically disadvantaged students in rural schools are not entirely clear.

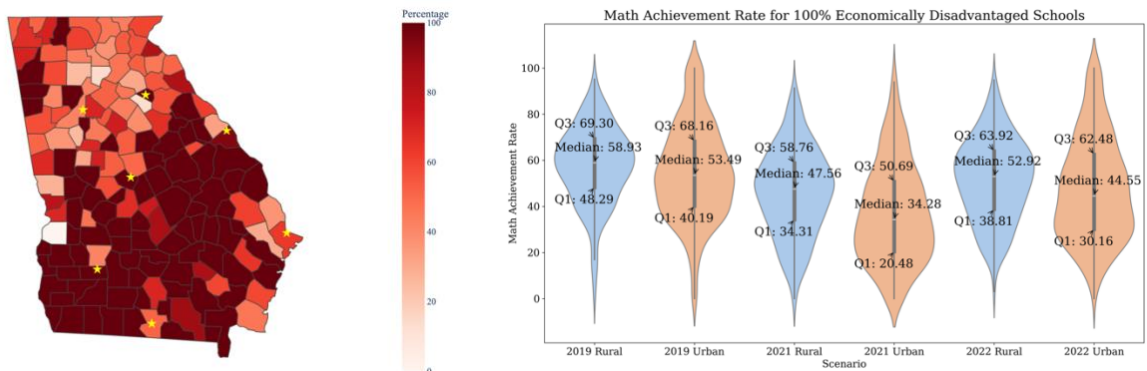

*Figure 11 (left). Visualization of Economically Disadvantaged Students Percentage in Georgia at County Levels*
*Figure 12 (right). Math Achievement Rate for 100% Economically Disadvantaged Schools*

In general, it is observed that the distribution of achievement rates for economically disadvantaged students in urban areas is skewed right. This means that there are more students performing at lower levels, and a smaller number of students performing at higher levels. On the other hand, the distribution of achievement rates for economically disadvantaged students in rural areas tends to be more symmetrical, with more students performing at average levels. This can be due to a variety of factors, including the supportive environment provided by rural communities, smaller class sizes, and a more personalized approach to education.

*Absenteeism*
Turning to absenteeism (Figure 13 A-C), this metric reflects the portion of students absent for certain ranges of days – less than 5 days, 6 to 15 days, and greater than 15 days, a critical indicator of chronic absenteeism (absent for more than 15 days is equivalent to absent for more than 8.33% of the academic year). Interestingly, this pattern mirrors that of the Economically Disadvantaged Percentage, with reasons ranging from limited transportation access, and health challenges, to unstable housing. Counties that have a relatively high percentage of students who are absent less than 5 days are Mitchell (92.2%), Hancock (85.6%), Wilkinson (82.2%), Terrel (81.7%), Union (77.1%), and Gwinnett (73.5%).

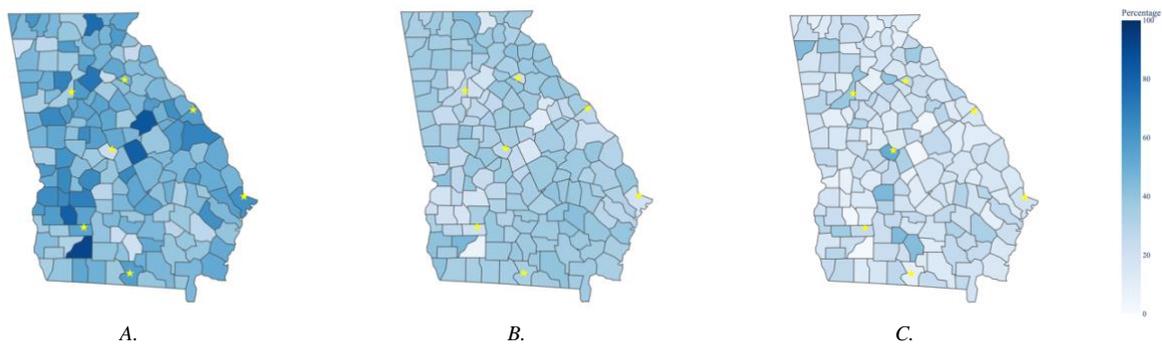

A.  B.  C.
*Figure 13 A. Visualization of Percentage of students who are absent for less than 5 days in Georgia at County Levels.*
*Figure 13 B. Visualization of Percentage of students who are absent for 6 – 15 days in Georgia at County Levels.*
*Figure 13 C. Visualization of Percentage of students who are absent for more than 15 days in Georgia at County Levels*

*Georgia Milestones End-of-Grade Assessment*

In this study, we analyze the performance of students in Metro counties using the Georgia Milestones Assessments results for Grade 3 and Grade 8 in the years 2019 and 2023 (Figures 14-17). For Grade 3, the data shows an increase in the percentage of distinguished learners in most of the metro counties, suggesting an improvement in high-performing students. However, there is also a rise in the percentage of beginning learners, indicating that students struggling academically are not showing the same level of progress. This divergence suggests a widening achievement gap, necessitating further investigation and intervention. In Grade 8, the percentage of distinguished learners has decreased significantly across all metro counties, indicating a reduction in high-performing students. Conversely, there is also a decrease in the percentage of beginning learners, which could imply that interventions for lower-performing students are having a positive effect. However, the overall decrease in top performers needs to be addressed to ensure equitable academic advancement. While there are positive trends in supporting lower-performing students, the widening achievement gap in Grade 3 and the decrease in high-performing Grade 8 students highlight areas that require further attention and action.

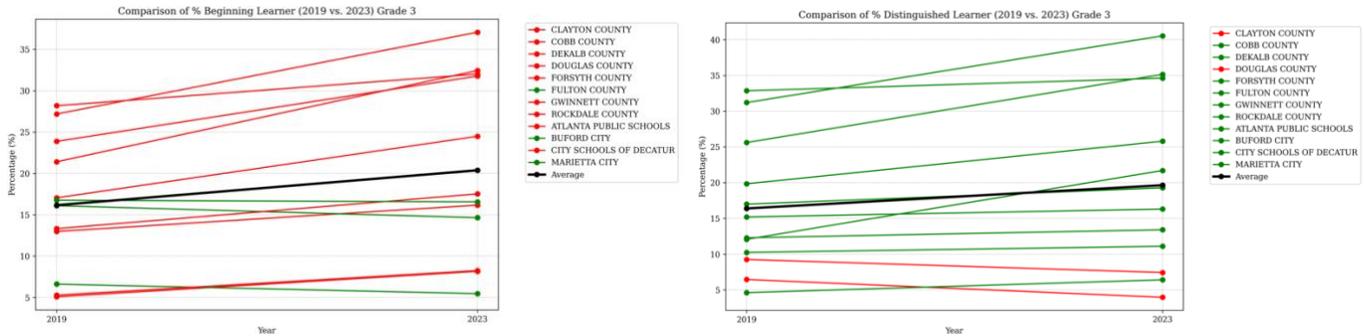

*Figure 14, 15. Beginning and Distinguished Learner Percentage for Grade 3 in Metro Counties in 2019 and 2023*

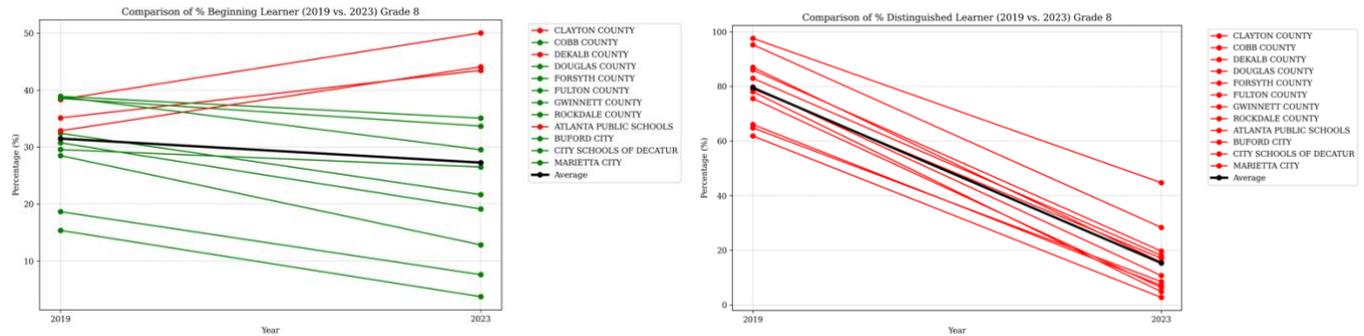

*Figure 16, 17. Beginning and Distinguished Learner Percentage for Grade 8 in Metro Counties in 2019 and 2023*

### 4.1.2. Machine Learning Models

*General Models*

To precisely predict outcomes and discern complex relationships within the data, some general machine learning models were employed. Evaluating the performance metrics, the Random Forest model notably stood out, explaining approximately 69.3% of the variability in the dataset. In contrast, Decision Trees, recognized for their interpretability, accounted for about 35.4% of the variability. The Stacking Regressor exhibited similar effectiveness as Random Forest, achieving a $R^2$ value of 0.695.

| *Models* | *MSE* | *RMSE* | *MAE* | $R^2$ |
|---|---|---|---|---|
| *Linear Regression* | 0.299 | 0.547 | 0.434 | 0.633 |
| *Decision Tree* | 0.527 | 0.726 | 0.458 | 0.354 |
| *Random Forest* | 0.251 | 0.501 | 0.385 | 0.693 |
| *MLP* | 0.277 | 0.526 | 0.415 | 0.660 |
| *SVM* | 0.259 | 0.509 | 0.394 | 0.682 |
| *KNN* | 0.305 | 0.552 | 0.424 | 0.626 |
| *Voting Regressor* | 0.257 | 0.507 | 0.399 | 0.685 |
| *Stacking Regressor* | 0.249 | 0.499 | 0.384 | 0.695 |
| *Bagging Regressor* | 0.275 | 0.525 | 0.395 | 0.662 |

*Table 4. General Machine Learning Models Performance*

*Deep Learning Models*

Deep learning models are known for their capacity to capture intricate patterns in large datasets. As we aimed to ensure accurate feature selection, three distinct models were devised and subsequently evaluated using the 10-Fold validation approach.

In Figure 5, Model #1 delivered an MSE of 113.900 and $R^2$ value of 0.793 denoting strong prediction accuracy and stability across validation folds. Model #2, however, surpassed Model #1 slightly, registering an MSE of 112.101 and an improved $R^2$ of 0.796 Conversely, Model #3 echoed the performance metrics of Model #1 with an MSE of 113.859 and an $R^2$ of 0.793. Among the models, Model #2 demonstrated a marginally superior ability to encapsulate the dataset's variability.

| Models | MSE | | RMSE | | MAE | |
|---|---|---|---|---|---|---|
| | Mean | Sd | Mean | Sd | Mean | Sd |
| Model #1 | 113.900 | 7.618 | 10.666 | 0.360 | 8.202 | 0.313 |
| | $R^2$ | | | | | |
| | Mean | | | Sd | | |
| | 0.793 | | | 0.018 | | |

| Models | MSE | | RMSE | | MAE | |
|---|---|---|---|---|---|---|
| | Mean | Sd | Mean | Sd | Mean | Sd |
| Model #2 | 112.101 | 10.706 | 10.576 | 0.507 | 8.068 | 0.315 |
| | $R^2$ | | | | | |
| | Mean | | | Sd | | |
| | 0.796 | | | 0.024 | | |

| Models | MSE | | RMSE | | MAE | |
|---|---|---|---|---|---|---|
| | Mean | Sd | Mean | Sd | Mean | Sd |
| Model #3 | 113.859 | 8.479 | 10.663 | 0.392 | 8.203 | 0.355 |
| | $R^2$ | | | | | |
| | Mean | | | Sd | | |
| | 0.793 | | | 0.018 | | |

*Table 5. Evaluation Metrics for Three Deep Learning Models*

Permutation Importance, an algorithm designed to compute the significance of each feature variable in a dataset relative to a given model, was implemented to determine each feature's importance. It assesses feature importance by examining how sensitive the model is to random shuffling of each feature's values. Essentially, the importance of a feature is gauged by observing the change in the model's prediction error after that feature's values have been permuted. If the model's error increases after shuffling a particular feature, it indicates that the model was reliant on that feature for its predictions, deeming it 'important.' Conversely, if the error remains unchanged post-permutation, it suggests that the feature wasn't influential in the model's predictions and is thus considered 'unimportant.' Figure 18 and 19 present the results from the Permutation Importance analysis and Table 6 presents the description of the results.

| *Directly Certified Percentage* | The percentage of students at a school who qualify for free and reduced lunch, is by far the most influential feature in all three models. A change in this feature is likely to have a significant impact on the model's prediction. The weight (or importance) of this feature is 262.1536 with a standard error of 36.2065, indicating the variability of this importance score repeating the permutation process multiple times. It is highly indicative of socioeconomic status and is a strong predictor in educational achievement models. A higher percentage typically corresponds to greater challenges in student achievement due to economic constraints. |
|---|---|
| *Grade* | The second most important feature with a weight of 81.4100 and a standard error of 10.4292, which is a lot less influential than the first but still significantly impacts the prediction. The Achievement rate in lower grades is higher than the achievement rate in higher grades, which may indicate the cumulative effect of educational challenges that become more apparent in higher grades. |
| *Teacher-Student Ratio* | With a weight of 46.3005, this feature's importance is about half that of Grade. Lower ratios typically correlate with better student outcomes due to more individualized attention. |
| *Demographic Features* | *Black, White, English Learners, Economically Disadvantaged,* and *Students with Disability* suggest that certain demographic groups might be influential in determining the outcome. The importance of these features might reflect underlying disparities or patterns in the data. |
| *Educational Features* | *4 years Bachelor's, 5 years Master's,* and *7 years Doctoral* seem to be proxies for teachers' attainment levels and they have varying degrees of importance. Higher education levels in teachers may correlate with better student achievement outcomes. |
| *Absenteeism* | Both *Absent for 6-15 Days Percentage* and *Absent Greater than 15 Days Percentage* are indicators of student absenteeism. It's notable that the former has a higher importance score, suggesting that being absent between 6 and 15 days is more impactful than being absent more than 15 days. Surprisingly, moderate absenteeism (6-15 days) appears to have a more substantial impact than severe absenteeism (>15 days), which may suggest a critical threshold for negative impacts on achievement. |

| Expenditure | Features like *Expenditure Instruction, Expenditure School Administration,* and *Expenditure Pupil Services* give insights into how different types of expenditures influence the outcome. The normalized instruction expenditure appears to be more influential than the others. Direct instructional expenditure has a more significant positive impact on student achievement than other types of school expenditure. |
|---|---|
| *Teacher-related Features* | *Teacher-Student Ratio, Average Teachers Annual Salary* point towards the importance of teacher-related factors. |

*Table 6. Description for Major Parameters Picked by the ML Model*

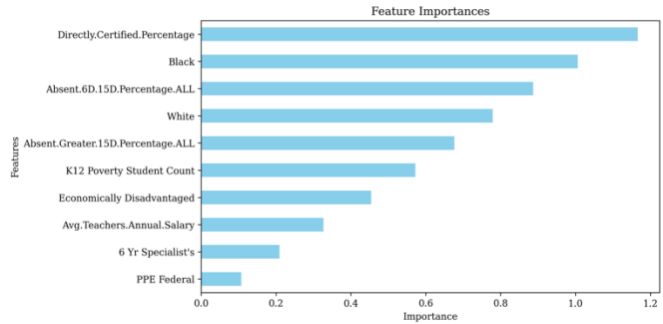

*Figure 18. Feature Importance for General Machine Learning Models*

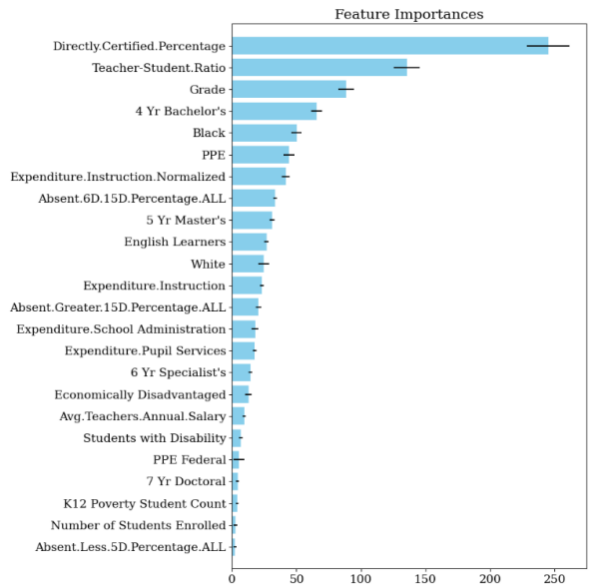

*Figure 19. Feature Importance for Deep Learning Model #2*

### 4.1.4. Schools Clustering

Clustering provides a methodological approach to systematically categorize schools into homogeneous groups with similar characteristics, enabling more refined analysis and clearer insights. The clustering technique used in this analysis is the K-Means algorithm, aiming to identify patterns and relationships that might not be immediately evident in a broader dataset. We partitioned the schools into four distinct clusters and the selection of features for clustering includes *Directly Certified Students Percentage*, *Average Teachers Annual Salary*, and *Economically Disadvantaged Students Percentage* (Figure 20). The selection of these particular features for clustering is logistically motivated and allows for a streamlined analysis of resource distribution and educational equity, as they provide a clear representation of a school's financial and socioeconomic status, which are crucial factors in educational resource allocation.

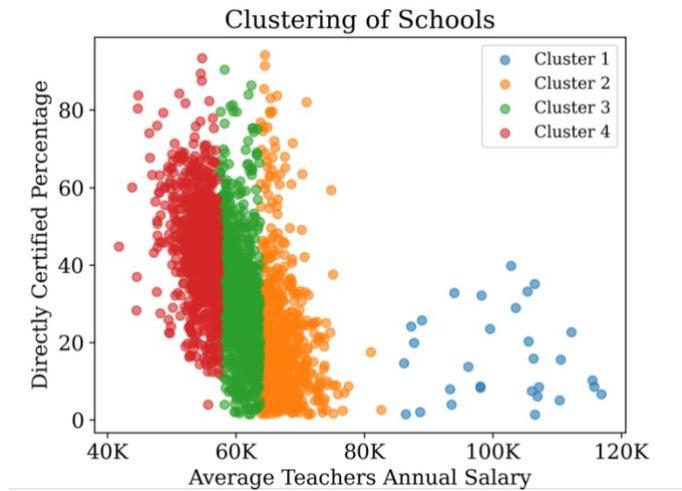

*Figure 20. kNN Clustering of Features Directly Certified Students Percentage and Average Teachers Annual Salary*

| Features | Cluster 1 | Cluster 2 | Cluster 3 | Cluster 4 |
|---|---|---|---|---|
| Achievement Math | 74.29 | 66.04 | 58.97 | 48.89 |
| Average Teachers Annual Salary | 1213.58 | 1286.64 | 1518.58 | 2157.51 |
| PPE Federal | 1213.58 | 1283.35 | 1506.37 | 2176.54 |
| Teacher-Student Ratio | 13.13 | 15.73 | 14.82 | 14.67 |
| Absent 6-15 Days Percentage | 45.78 | 34.05 | 36.60 | 37.25 |
| Absent Greater than 15 Days Percentage | 21.67 | 18.96 | 20.08 | 23.28 |
| White and Black Students Disparity | 22.47 | 23.45 | 25.53 | 23.54 |
| Economically Disadvantaged Students Disparity | 17.37 | 11.60 | 9.00 | 2.84 |
| White Percentage | 34.55 | 38.56 | 43.02 | 37.80 |
| Black Percentage | 31.40 | 33.30 | 32.83 | 43.27 |
| Economically Disadvantaged Percentage | 25.84 | 47.28 | 60.03 | 82.99 |
| English Learners Percentage | 15.43 | 10.62 | 9.90 | 6.94 |
| Students with Disability Percentage | 14.76 | 13.33 | 14.50 | 15.01 |
| Expenditure Instruction | 9356.74 | 7137.05 | 6825.03 | 6277.92 |
| 6 Years Specialist's | 0.01 | 0.01 | 0.01 | 0.01 |
| 7 Years Doctoral | 0.00 | 0.00 | 0.00 | 0.00 |
| Directly Certified Percentage | 16.17 | 20.98 | 31.67 | 44.26 |

*Table 7. kNN Clusters Statistics*

The violin plot visualization compares high performing and low performing clusters across various metrics. High performing clusters show a higher median with a broader distribution in Math Achievement rate, indicating greater variance in scores (Figure 21). The low performing cluster displays a wider distribution in Directly Certified Percentage and a slightly higher median for Economically Disadvantaged students percentage. The Average Teachers Annual Salary reveals higher salaries in the high performing cluster with a narrow distribution, implying more consistency in pay. When considering the achievement gap, the Economically Achievement Gap has a wide distribution for low performing clusters, indicating significant disparity. In contrast, the distribution for Racial Disparity is relatively similar between the two clusters. Finally, the Teacher-Student Ratio in high performing clusters shows a narrow range, suggesting more uniformity in teacher-student ratios, whereas the low performing cluster indicates a wider spread, reflecting greater variation in ratios. Overall, high performing clusters tend to have higher achievement in math, higher average teacher salaries, and a more consistent teacher-student ratio, while low performing clusters exhibit more variation in certification, economic disadvantage, and disparities.

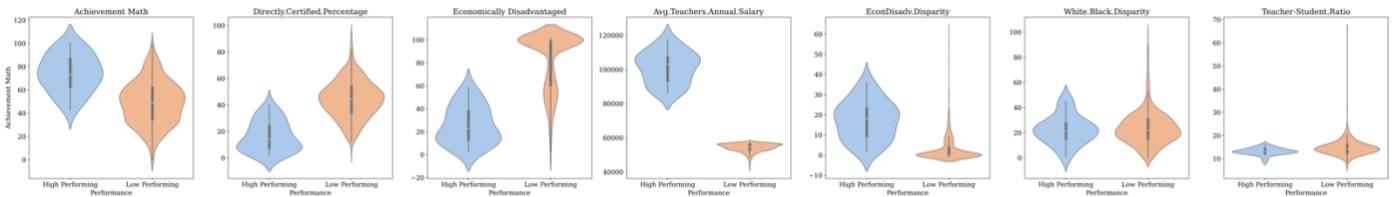

*Figure 21. Violin Plots for High and Low Performing Clusters*

## 4.2. Achievement Gap
### 4.2.1. Racial Achievement Gap
Our analysis concentrates on the achievement gap between White and Black students due to the limited availability of comprehensive educational resource data for each minority group. We acknowledge the presence of other ethnicities; however, the dataset provides an insufficient number of data points for these groups, resulting in high variability and reduced reliability in the findings. Figure 22 illustrates the correlation between schools' Math Achievement rate and the Achievement Gap between White and Black students with each dot representing each's school's total students count on the same scale. The data points appear to be widely dispersed, indicating variation in the relationship between the two variables; as Math Achievement rates increase, there's a notable range in the White-Black Disparity. Furthermore, the color gradient, indicating the percentage of Black students, transitions from yellow (0%) to deep purple (100%). At lower Math Achievement levels, there is a clustering of points reflecting smaller White-Black Achievement Gaps, typically represented by lighter shades indicating a higher prevalence of Black students. In contrast, at higher Math Achievement levels, the extent of the White-Black Achievement Gap broadens, and these data points are frequently marked by darker hues, suggesting a reduced representation of Black students. In summary, the plot suggests that areas with lower math achievement scores often correspond with a lower White-Black achievement disparity and a higher percentage of Black students. Conversely, as math achievement scores increase, the White-Black achievement disparity presents a wider range and often accompanied by a lower percentage of Black students.

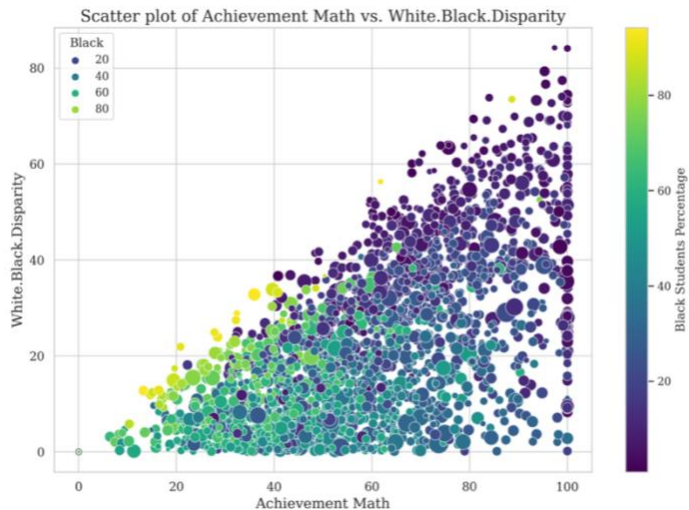

*Figure 22. Scatter Plot of Students Math Achievement vs. Racial Achievement Gap*

*Model Insights on Racial Achievement Gap*
Our model produced a moderate fit and result for the racial achievement gap. Figure 23 represents feature importance in predicting the racial achievement gap between white and black students. The most influential feature is Directly Certified Percentage, followed by PPE, suggesting that percentage of students who are qualified free lunch and per-pupil expenditure significantly affect the gap. The other features, such as English Learners and Teacher-Student Ratio, also play roles, but to a lesser extent. In terms of model accuracy, a 10-fold cross-validation method was employed, yielding the following statistics: a Mean Squared Error (MSE) of 118.266 (SD = 12.803), a Root Mean Squared Error (RMSE) of 10.859 (SD = 0.590), and a Mean Absolute Error (MAE) of 8.485 (SD = 0.340). The model's average $R^2$ value stood at 0.574 (SD = 0.041), indicating a moderate degree of variance explanation. Interestingly, a trend was observed where higher student achievement scores are associated with a larger achievement gap. This warrants a deeper analysis, especially considering the notable cluster of schools with high achievement scores, which could provide further insights into the dynamics of educational disparities.

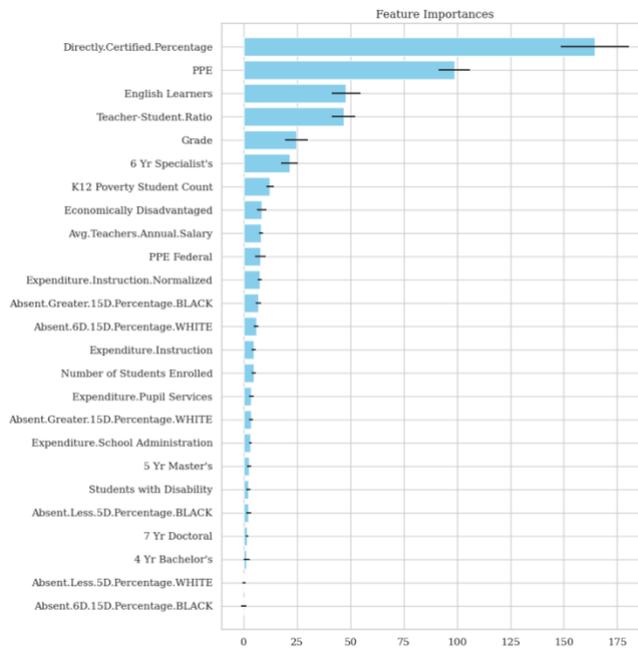

*Figure 23. Feature Importance Plot for the Deep Learning Model of predicting Racial Achievement Gap*

| Models | MSE | | RMSE | | MAE | |
|---|---|---|---|---|---|---|
| | *Mean* | *Sd* | *Mean* | *Sd* | *Mean* | *Sd* |
| *Model #3* | 118.266 | 12.803 | 10.859 | 0.590 | 8.485 | 0.340 |
| | $R^2$ | | | | | |
| | *Mean* | | | *Sd* | | |
| | 0.574 | | | 0.041 | | |

*Table 8. Evaluation Metrics for the Model of predicting Racial Achievement Gap*

We then used Individual Conditional Expectation (ICE) plot to gain a high-level overview of feature effects, where it visualizes the predicted outcome for an individual data point as a function of a specific feature, while keeping all other features constant. Each line in an ICE plot represents the model's prediction for a single instance as the feature of interest changes over its range. We have the y-axis as White and Black students' Achievement Rate Disparity and the x-axis as the range of the selected feature of interest. Figure 24 shows that for all 10 ICE plots, all lines (representing different instances) follow a similar trajectory, which indicates a consistent relationship across the dataset. Features, such as Directly Certified Percentage, PPE, English Learners Percentage, and Teacher-Student Ratio, displays a negative correlation with the Achievement gap between White and Black students. Features, such as 6-years Specialists Count and PPE Federal, have a positive correlation with the racial achievement gap, indicating that by increasing their values will also increase the racial achievement gap.

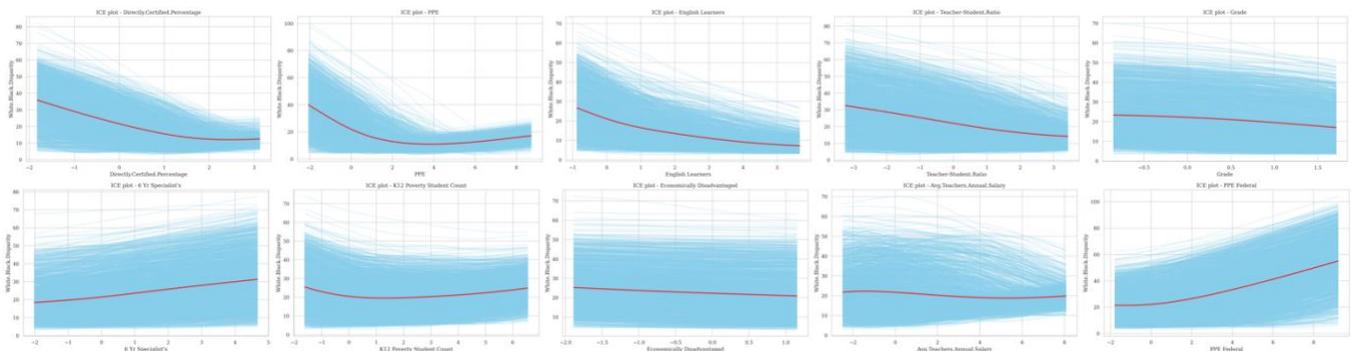

*Figure 24. ICE Plots for Dependency of the Racial Achievement Gap with Top Features Selected*

#### 4.2.2. Economical Achievement Gap
In addition to the achievement gap in racial groups, we are interested in studying the relationship within each school between Economically disadvantaged students and all students. Figure visualizes the relationship between a school's Math Achievement rate and Economical Disparity, differentiated by the percentage of students categorized as Economically Disadvantaged. Data points extend from the bottom left to the top right of the graph, indicating a positive correlation between the two variables. The color gradient, which represents the percentage of economically disadvantaged students, transitions from yellow (0%) to deep purple (100%). At lower levels of Achievement Math, there's a predominance of higher percentages of economically disadvantaged students, as evident by the lighter shades of yellow. In contrast, as the levels of Achievement Math rise, the percentage of economically disadvantaged students seems to decrease, marked by a shift in color from yellow to green, and eventually to purple. This observation suggests that schools or regions with higher mathematical achievement tend to have a reduced percentage of economically disadvantaged students and a higher economically achievement gap incurred.

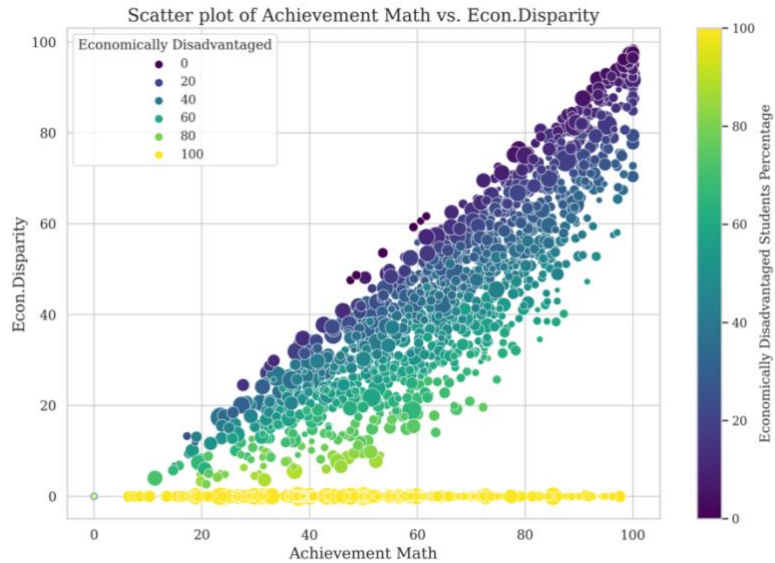

*Figure 25. Scatter Plot of Students Math Achievement vs. Economical Achievement Gap*

Our model produced a better fit and result for the economic achievement gap. Figure 32 represents feature importance in predicting the economical achievement gap. Similar to the racial achievement gap predictions, the most influential feature is Directly Certified Percentage, but followed by PPE Federal, instead of PPE. The rest of features play a negligent rote compared to the Directly Certified Percentage. In a 10-fold validation, the model achieved a Mean Squared Error (MSE) of 114.518 with a standard deviation (Sd) of 15.910, a Root Mean Squared Error (RMSE) of 10.676 with an Sd of 0.737, and a Mean Absolute Error (MAE) of 7.507 with an Sd of 0.593. The average $R^2$ across the folds was 0.860 with an Sd of 0.019.

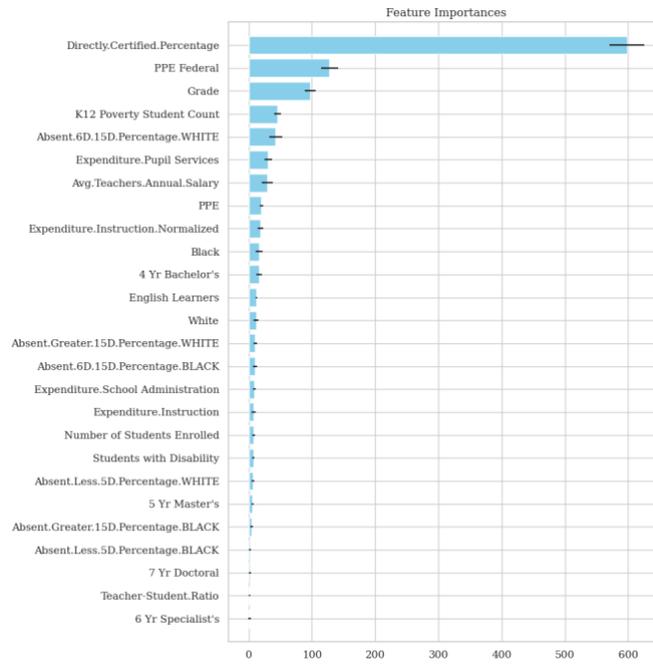

*Figure 26. Feature Importance Plot for the Deep Learning Model of predicting Economical Achievement Gap*

| Models | MSE | | RMSE | | MAE | |
|---|---|---|---|---|---|---|
| | *Mean* | *Sd* | *Mean* | *Sd* | *Mean* | *Sd* |
| *Model #3* | 114.518 | 15.910 | 10.676 | 0.737 | 7.507 | 0.593 |
| | $R^2$ | | | | | |
| | *Mean* | | | *Sd* | | |
| | 0.860 | | | 0.019 | | |

*Table 9. Evaluation Metrics for the Model of predicting Economical Achievement Gap*

The economic achievement gap is the difference between the achievement rate for all students and the achievement rate for ece economically disadvantaged students. The ICE plots for the economic achievement gap reveals that with a higher Directly Certified Percentage, the economic achievement gap shrinks. PPE federal and high expenditure on instruction also have a negative correlation with the gap, but the slope is flatter.

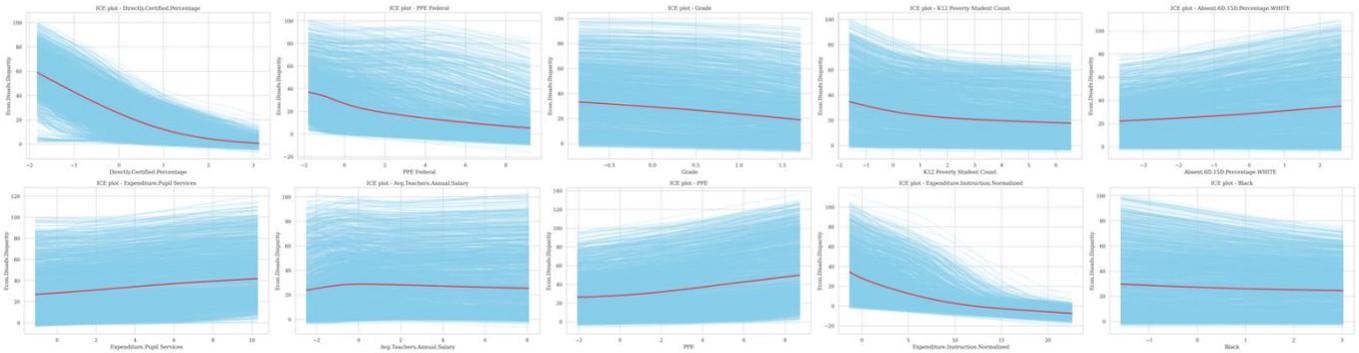

*Figure 27. ICE Plots for Dependency of the Economical Achievement Gap with Top Features Selected*

## 5. Discussion

The application of machine learning models has provided valuable insights into the factors influencing student achievement. The Random Forest model, in particular, has demonstrated its efficacy in explaining a significant portion of the variability in the dataset. The importance of features such as Directly Certified Percentage and Grade in these models underscores the critical role of socio-economic status and educational level in student achievement. These insights provide a data-driven basis for targeted interventions and resource allocation, ensuring that support is directed where it is needed most.

**5.1. Overview of Achievement Rates**

The comprehensive analysis of student achievement rates conducted in this research has revealed a nuanced and complex landscape of educational performance across different demographics, regions, and subjects. The period between 2019 and 2021, marked by the COVID-19 pandemic, has seen a significant decline in proficiency across both English and Math, a trend that raises urgent questions about the state of education and the long-term impacts of such disruptions.

The noticeable decline in achievement rates across English and Math is a critical concern, reflecting potential systemic issues in education delivery and student engagement. The mean achievement rate in English saw a reduction from 64.40% to 51.08%, and in Math from 64.58% to 47.23%. These are not marginal decreases; they represent a substantial drop in proficiency, with potentially far-reaching consequences for the students affected. The narrowing of the interquartile range in both subjects further indicates a contraction in score distribution, suggesting that fewer students are reaching the higher achievement quartiles. This phenomenon could imply a loss of high-achieving students or a general shift of the student population towards lower achievement levels, each scenario having distinct implications for educational policy and intervention strategies.

The data highlights disparities in achievement rates among different demographic groups. Economically disadvantaged students and Black students are particularly affected, with their mean achievement rates significantly lower than their peers. This disparity is an indicator of the systemic inequalities that pervade the educational system, reflecting socio-economic and racial factors that influence student performance. Addressing these disparities is not just a matter of educational justice; it is a societal imperative, as education is a key driver of social mobility and economic stability. The differences in achievement rates between urban and rural settings, as well as across various counties, further emphasize the role of geographical and socio-economic factors. The slight edge that rural counties have over urban counties in 2021, despite having lower resources, raises questions about the resilience of different educational systems and the factors that contribute to student success in varied settings.

**5.2. Pivotal Role of Socio-Economic Status**

Directly Certified Percentage

Georgia education landscape presents a multifaceted challenge with the presence of a significant percentage of economically disadvantaged students in its public schools. The Directly Certified Percentage, reflecting the proportion of students eligible for free lunch, has been consistently highlighted by multiple machine learning models as the most influential factor in student achievement rates across various demographics and regions. This underscores the profound impact of socio-economic status on educational outcomes, necessitating targeted interventions and policy reforms to mitigate its effects and promote equity. The consistent prominence of this factor across diverse models reinforces the urgency of addressing socio-economic disparities in education, ensuring that all students have equal opportunities to succeed, regardless of their financial background.

Further on the socio-economic disparities, economic disadvantage, a term often tethered to limited access to resources, invariably leads to lower academic achievements and graduation rates. This discussion delves into the intricate patterns observed in the achievement rates of students in Georgia public schools, focusing on those who are 100% economically disadvantaged. The results indicate a significant decline in achievement rates for economically disadvantaged students in Georgia between 2019 and 2021, underscoring the prevalent impact of economic status on educational outcomes. This decline is particularly notable in light of the COVID-19 pandemic, which has intensified existing inequalities and posed additional challenges to students already at a disadvantage. However, a striking revelation is the difference in performance between urban and rural schools. Contrary to many assumptions, our analysis revealed that economically disadvantaged students in rural areas outperformed their urban counterparts. In 2019, urban schools might have provided better support systems, leading to higher achievement rates. However, the pandemic seems to have shifted this dynamic, with rural schools demonstrating a surprising resilience. The underlying reasons for this shift remain unclear. This shift could be attributed to the close-knit communities and stronger support systems often found in rural areas, which may have played a crucial role in mitigating the educational disruptions caused by the pandemic.

Illustratively, the distribution of achievement rates for economically disadvantaged students in urban areas skews right, indicating a larger proportion of students scoring below average. This skew can be attributed to the challenges that economically disadvantaged students in urban areas may face, such as limited access to resources, higher crime rates, and social and economic instability, which inadvertently create barriers to academic success. Economically disadvantaged students in these areas grapple not only with their economic status but also with the associated urban challenges. Conversely, the achievement rate distribution for economically disadvantaged students in rural areas aligns more with a normal distribution. A significant portion of students achieve average scores, indicating a balance in performance. Rural communities often offer a support system, potentially augmented by smaller class sizes and a tailored education approach, all of which can enhance learning experiences. Moreover, the superior performance of economically disadvantaged students in rural areas, as opposed to their urban counterparts, might also be steering this normal distribution, indicating a broader spectrum of students achieving commendable scores.

**5.3. Faculty Resources**

Teacher salaries and the teacher-student ratio are crucial factors influencing student achievement, as revealed by our machine learning models. Competitive teacher salaries are essential for attracting and retaining high-quality educators, which directly correlates with improved student outcomes. Investing in teacher quality through fair compensation not only acknowledges the value of educators but also contributes to a more stable and experienced teaching workforce, fostering a positive learning environment. The teacher-student ratio holds equal significance, with lower ratios often leading to more personalized instruction and better academic achievement. A manageable teacher-student ratio ensures that educators can give individualized attention to each student, addressing specific needs and adapting teaching methods accordingly. This is particularly crucial for students who may require additional support, ensuring that no student is left behind due to a lack of attention or resources. Therefore, a balanced approach that addresses both teacher remuneration and class size is essential. Policymakers must prioritize initiatives that not only offer competitive salaries to attract the best teaching talent but also ensure optimal class sizes for effective teaching and learning. This dual focus is vital for enhancing educational outcomes and promoting equity across different student groups and regions.

### 5.4. Racial Achievement Gap

The features displaying a negative correlation with the racial achievement gap, such as Directly Certified Percentage, Per-Pupil Expenditure (PPE), English Learners Percentage, and Teacher-Student Ratio, suggest that as the values of these features increase, the achievement gap tends to decrease. A higher Directly Certified Percentage may indicate more resources and support directed towards economically disadvantaged students, which could disproportionately benefit Black students if they are overrepresented in this group, thereby reducing the achievement gap. Similarly, increased PPE and lower Teacher-Student Ratios can lead to more resources and attention per student, potentially mitigating disparities. The English Learners Percentage might play a role as schools with higher percentages of English learners may receive additional support and resources, indirectly benefiting all student groups and narrowing the racial achievement gap.

Conversely, features such as 6-years Specialists Count and PPE Federal showing a positive correlation with the racial achievement gap imply that as these values increase, the gap widens. A higher count of specialists with six years of experience may indicate a distribution of experienced educators that is not equitably benefiting Black students, potentially due to systemic issues in school assignments or teacher placements. Additionally, an increase in federal funding per student (PPE Federal) not leading to a reduction in the racial achievement gap may suggest that these funds are not being utilized effectively to address the specific needs of Black students or there are other prevailing issues that the additional funding is not able to overcome.

### 5.5. Economic Achievement Gap

The economic achievement gap, defined as the discrepancy in achievement rates between all students and economically disadvantaged students, is a crucial metric for assessing equity in education. The Individual Conditional Expectation (ICE) plots utilized in our analysis provide valuable insights into the factors influencing this gap, particularly highlighting the role of the Directly Certified Percentage, Per-Pupil Expenditure (PPE) Federal, and expenditure on instruction. The negative correlation observed between the Directly Certified Percentage and the economic achievement gap suggests that as the percentage of students eligible for free lunch (and thus likely to be economically disadvantaged) increases, the achievement gap tends to decrease. This could be indicative of targeted interventions and resources being effectively allocated to support economically disadvantaged students in schools with higher Directly Certified Percentages. It underscores the importance of providing additional support to students in need, ensuring that they have access to the necessary resources and opportunities to succeed academically. PPE Federal and expenditure on instruction also exhibit a negative correlation with the economic achievement gap, albeit with a flatter slope compared to the Directly Certified Percentage. This implies that while increased federal funding per student and higher investment in instruction contribute to narrowing the achievement gap, their impact might be more gradual and less pronounced than that of direct support to economically disadvantaged students. It is crucial to ensure that the additional funds and resources are being utilized effectively to address the specific needs of these students, promoting equity and fostering an inclusive learning environment. The analysis of the economic achievement gap through ICE plots reveals the critical role of socio-economic factors and resource allocation in influencing educational equity.

### 5.6. Potential Methods

Achieving equity and access to high-quality education requires comprehensive measures that address factors such as resource allocation, teacher recruitment and training. (1) Address absenteeism: Implement policies and programs to reduce chronic absenteeism, such as improving access to transportation and healthcare, and addressing unstable housing situations. (2) Increase teacher and administrator salaries: Improve teacher and administrator salaries to attract and retain high-quality educators, which can positively impact student achievement. (3) Address the achievement gap: Develop programs and initiatives that address the socio-economic status of students, particularly those in economically disadvantaged areas. This may include increasing funding for schools in disadvantaged areas, providing additional resources and support to students and families, and expanding access to early childhood education. (4) Reduce class size: Increase efforts to reduce class sizes in schools. (5) Encourage diversity: Foster a diverse and inclusive learning environment by promoting diversity in the student population, hiring practices, and curriculum.

# Appendix

## A. Grade Levels

*Directly Certified Percentage*, *White Students Percentage*, and *Black Students Percentage* serves as the most important features for all grade classifications. The feature importance for all three grade classifications has high similarity, we only included Figure 15 of high schools for clarity.

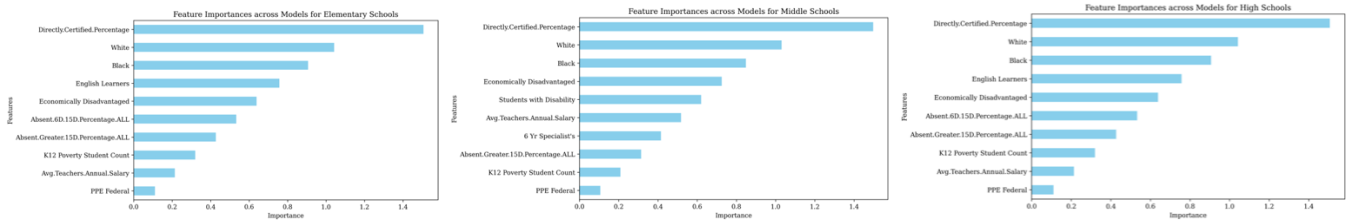

*Figure 28. Feature Importance for Elementary, Middle, and High School Achievement Rate Prediction*

## B. Deep Learning Models Feature Importance

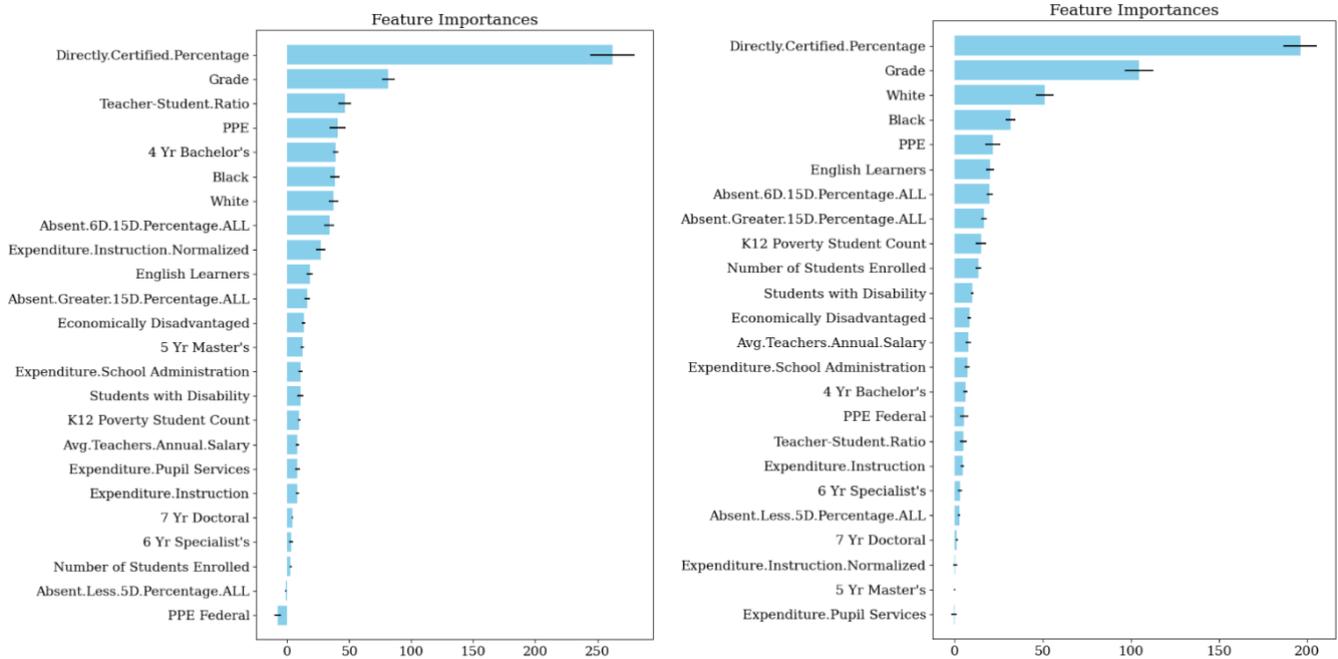

*Figure 29. Feature Importance for Deep Learning Model #1, 3*

## C. Partial Dependency

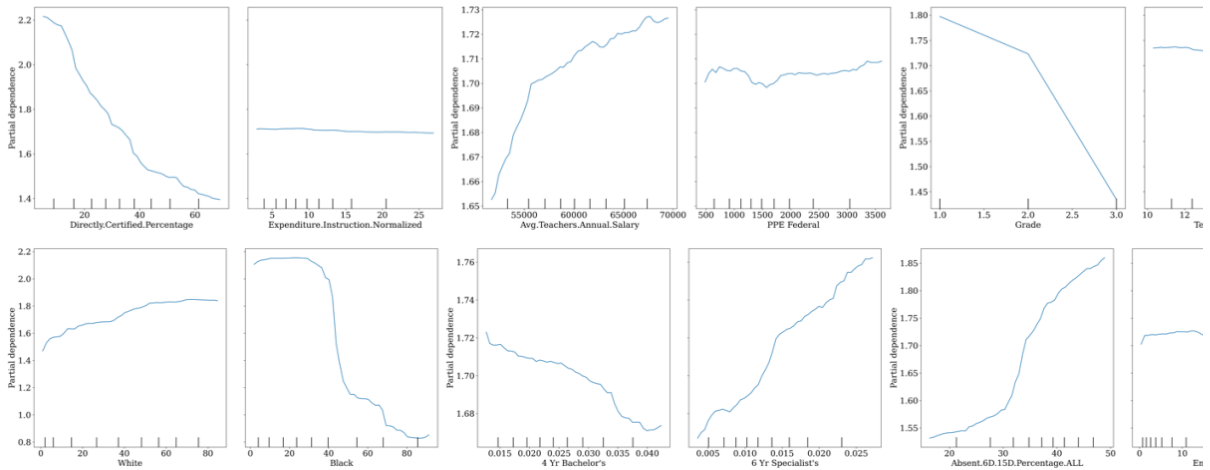

*Figure 30. Partial Dependency Visualizations*

## D. KNN on the Directly Certified Percentage and the Economically Disadvantaged Students Percentage

The KNN clustering of schools based on the Directly Certified Percentage and Average Teachers Annual Salaries yielded four distinctive clusters. Cluster 1 comprises schools with the highest math achievement, the lowest teacher-student ratio, and modest direct certification rates. In contrast, Cluster 4 represents schools with the lowest math performance, the highest Directly Certified Percentage, and the lowest expenditure on instruction. Interestingly, while Cluster 2 and 3 have mid-range math achievements, Cluster 2 exhibits a significantly higher percentage of economically disadvantaged students than Cluster 3. Across the clusters, the variation in disparities between White and Black students remains minimal, although economic disparities decrease markedly from Cluster 1 to 4. The information suggests a correlation between economic factors and academic performance, with schools in economically challenged settings having reduced academic outcomes.

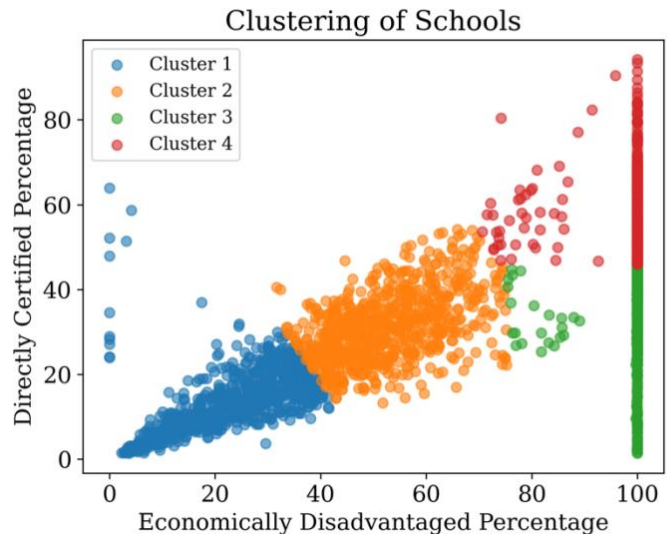

*Figure 31. kNN Clustering of Features Directly Certified Students Percentage and Average Teachers Annual Salary*

| Features | Cluster 1 | Cluster 2 | Cluster 3 | Cluster 4 |
|---|---|---|---|---|
| Achievement Math | 75.08 | 57.56 | 52.06 | 41.18 |
| Average Teachers Annual Salary | 65226.31 | 60542.28 | 59454.94 | 56407.82 |
| PPE Federal | 1018.57 | 1406.27 | 2059.17 | 2453.58 |
| Teacher-Student Ratio | 15.92 | 15.03 | 14.85 | 13.88 |
| Absent 6-15 Days Percentage | 38.05 | 36.22 | 34.45 | 35.52 |
| Absent Greater than 15 Days Percentage | 17.17 | 20.68 | 21.33 | 25.29 |
| White and Black Students Disparity | 24.42 | 24.82 | 24.65 | 23.05 |
| *Economically Disadvantaged Students Disparity* | 18.35 | 8.69 | 0.09 | 0.20 |
| *White Percentage* | 50.62 | 40.38 | 39.04 | 26.61 |
| *Black Percentage* | 21.05 | 34.03 | 37.58 | 58.09 |
| *Economically Disadvantaged Percentage* | 22.91 | 52.42 | 99.14 | 98.37 |
| *English Learners Percentage* | 8.73 | 11.21 | 9.74 | 6.64 |
| *Students with Disability Percentage* | 13.44 | 14.95 | 14.49 | 14.58 |
| *Expenditure Instruction* | 6881.21 | 6853.87 | 6492.61 | 6757.73 |
| *6 Years Specialist's* | 0.01 | 0.01 | 0.01 | 0.01 |
| *7 Years Doctoral* | 0.00 | 0.00 | 0.00 | 0.00 |
| *Directly Certified Percentage* | 13.73 | 31.15 | 34.35 | 57.63 |

*Table 5.*

The kNN clustering of schools based on Directly Certified Percentage and Economically Disadvantaged Percentage again reveals similar distinct groups. Cluster 1 schools, with the highest math achievements (75.08) and moderately high teacher salaries, exhibit lower percentages of economically disadvantaged (22.91%) and directly certified students (13.73%). Cluster 2, while having a math achievement of 57.56, sees a substantial rise in economically disadvantaged (52.42%) and directly certified students (31.15%). Cluster 3 is characterized by almost entirely economically disadvantaged students (99.14%) with a math achievement of 52.06 and the least expenditure on instruction. Cluster 4, with the lowest math achievement (41.18), reports the highest directly certified students (57.63%), predominantly Black student population (58.09%), and the highest federal per-pupil expenditure. This clustering underscores the relationship between economic disadvantage, direct certification, and academic achievement.